\documentclass{aa}  

\usepackage{graphicx}
\usepackage{txfonts}
\usepackage[]{hyperref}
\newcommand{\revised}[1]{#1}
\newcommand{\revisedtwo}[1]{#1}

\begin{document}

   \title{White dwarf deflagrations for Type Iax supernovae: Polarisation signatures from the explosion and companion interaction}

    \titlerunning{White dwarf deflagrations for Type Iax supernovae: Polarisation signatures}

   \author{M.~Bulla
          \inst{1}
          \and
          Z.-W. Liu\inst{2,3,4}
          \and
          F. K. R\"{o}pke\inst{5,6}
          \and
          S. A. Sim\inst{7}
          \and
          M. Fink\inst{8}
          \and
          M. Kromer\inst{5,6}
          \and
          R. Pakmor\inst{9}
          \and
          I.~R.~Seitenzahl\inst{10}
          }

   \institute{Nordita, KTH Royal Institute of Technology and Stockholm University, Roslagstullsbacken 23, SE-106 91 Stockholm, Sweden\\
              \email{mattia.bulla@fysik.su.se}
         \and Yunnan Observatories, Chinese Academy of Sciences (CAS), Kunming 650216, P.R. China
         \and Key Laboratory for the Structure and Evolution of Celestial Objects, CAS, Kunming 650216, P.R. China
         \and Center for Astronomical Mega-Science, CAS, Beijing, 100012, P. R. China
         \and 
         Zentrum f{\"u}r Astronomie der Universit{\"a}t Heidelberg, Institut f{\"u}r Theoretische Astrophysik, Philosophenweg 12, 69120 Heidelberg, Germany
         \and
         Heidelberger Institut f{\"u}r Theoretische Studien, Schloss-Wolfsbrunnenweg 35, 69118 Heidelberg, Germany
         \and
         Astrophysics Research Centre, School of Mathematics and Physics, Queen's University Belfast, BT7 1NN, UK
         \and Institut f\"ur Theoretische Physik und Astrophysik, Universit\"at W\"urzburg, Emil-Fischer-Stra{\ss}e 31, D-97074, W\"urzburg, Germany
         \and Max-Planck-Institut f\"{u}r Astrophysik, Karl-Schwarzschild-Str. 1, D-85748, Garching, Germany
         \and
         School of Science, University of New South Wales, Australian Defence Force Academy, Northcott Drive, Canberra, ACT 2600, Australia}

   \date{Received; accepted}
    
    \abstract{Growing evidence suggests that Type Iax supernovae might be the result of thermonuclear deflagrations of Chandrasekhar-mass white dwarfs in binary systems. We carry out Monte Carlo radiative transfer simulations and predict spectropolarimetric features originating from the supernova explosion and subsequent ejecta interaction with the companion star. Specifically, we calculate viewing-angle dependent flux and polarisation spectra for a 3D model simulating the deflagration of a Chandrasekhar-mass white dwarf and, for a second model, simulating the ejecta interaction with a main-sequence star. We find that the intrinsic signal is weakly polarised and only mildly viewing-angle dependent, owing to the overall spherical symmetry of the explosion and the depolarising contribution of iron-group elements dominating the ejecta composition. The interaction with the companion star carves out a cavity in the ejecta and produces a detectable, but modest signal that is significant only at relatively blue wavelengths ($\lesssim$~5000~\AA). In particular, increasingly fainter and redder spectra are predicted for observer orientations further from the cavity, while a modest polarisation signal $P\sim0.2$~per cent is found at blue wavelengths for orientations 30$^\circ$ and 45$^\circ$ away from the cavity. We find a reasonable agreement between \revised{the interaction model viewed from these orientations} and spectropolarimetric data of SN~2005hk and interpret the maximum-light polarisation signal seen at blue wavelengths for this event as a possible signature of the ejecta--companion interaction. \revised{We encourage further polarimetric observations of SNe Iax to test whether our results can be extended and generalised to the whole SN Iax class.} }


\maketitle

\section{Introduction}
\label{sec:introduction}

Type Ia supernovae (SNe Ia) are widely believed to stem from thermonuclear explosions of carbon--oxygen white dwarfs in binary systems, but the exact conditions leading to the runaway are debated (see e.g. \citealt{Livio2018} for a recent review). Available scenarios differ depending on whether the white dwarf explodes near or below the Chandrasekhar mass ($M_\mathrm{ch}$), and on whether the binary companion star is a non-degenerate star in a `single-degenerate' system \citep{whelan1973a} or another white dwarf in a `double-degenerate' system \citep{iben1984a,webbink1984a}. 
From observables around peak luminosity alone, it is difficult to distinguish between the scenarios \citep[e.g.][]{roepke2012a}, but nuclear physics effects lead to imprints on the chemical structure of the ejecta \citep[e.g.][]{seitenzahl2013b,Yamaguchi2015,floers2018,floers2019}.

A striking difference between the single-degenerate progenitor scenario and white dwarf mergers is the presence of a close companion in the former. The SN ejecta significantly interact with this companion star after the explosion, stripping some hydrogen/helium-rich material from the companion surface and creating a cavity in the SN ejecta, but the companion survives as a bound stellar object
(e.g. \citealt{wheeler1973a, Marietta2000a, Pakmor2008a, Liu2012a, Pan2012a, Bauer2019}). This has a number of implications. If SNe~Ia result from single-degenerate progenitors, it should be possible to detect the surviving companions in their remnants\footnote{We note that also for the double-degenerate progenitor model in the so-called D6 scenario, surviving white dwarf companions are expected. \cite{shen2018a} propose candidates in the Gaia sample.} (see \citealt{RuizLapuente2018} for a recent review and \citealt{Li2019} for a recent search). Although for the Tycho remnant such a detection has been claimed (\citealt{ruiz-lapuente2004a}, but see \citealt{Kerzendorf2018}) other campaigns could not identify suitable candidates \citep[e.g.][]{schaefer2012a,Shappee2013b}. Another implication is the enrichment of the SN ejecta with material stripped off from the companion by the SN blast wave. Corresponding hydrogen/helium lines are expected to be present in late-time spectra of SNe Ia (e.g. \citealt{mattila2005a,lundqvist2013,Botyanszki2018}), but no hydrogen/helium line has been detected yet in late-time observations (e.g. \citealt{leonard2007a,shappee2013a,lundqvist2015a,maguire2016a,Tucker2019a, Tucker2019b}) except in two fast-declining, fsub-luminous events \citep{Prieto2019,Kollmeier2019}.

The presence of a cavity during the ejecta--companion interaction introduces a large-scale asymmetry in the SN ejecta, thus leading to a polarisation signal for favourable viewing angles.
Polarisation predictions were computed by \citet{kasen2004a} for a simplified structure and composition of the SN ejecta. No continuum polarisation was found when looking down the cavity (viewing angle 0 degrees) because the ejecta are symmetric in projection, while non-zero signals were predicted for views away from the cavity with a maximum at 90~degrees.  

In this pilot study, we derive the observational imprints of the SN interaction with a companion star in a $M_\mathrm{ch}$ explosion model. Pure deflagrations in $M_\mathrm{ch}$ carbon--oxygen white dwarfs have been shown to reproduce key properties of so-called `2002cx-like' SNe \citep{phillips2007a, jordan2012b, kromer2013a, fink2014a} and perhaps of the more extended class of Type Iax SNe \citep[SNe~Iax, see][for a recent review]{jha2017a}.
One event of this class, SN~2012Z, represents the only case in which a progenitor system of a thermonuclear SN has been identified in pre-explosion images \citep{mccully2014a}, suggesting a possible connection between SNe~Iax and $M_\mathrm{ch}$ white dwarfs exploding in single-degenerate systems. Although the progenitor in SN~2012Z was identified as a He star, in this paper we model the interaction with a main-sequence (MS) star and address the impact of this choice on our conclusions. 

\revised{We note that SNe~Iax show large diversity and it is yet not clear whether they result from one single progenitor channel. In particular, a layered ejecta structure has been suggested to explain SNe Iax \citep{Stritzinger2015a,barna2017a,barna2018a}, a stratification that is hard to reconcile with the highly mixed ejecta produced in pure deflagration models. \revisedtwo{The layered structure inferred for SN 2012Z, together with the lack of oxygen and intermediate-mass elements (IME) in the inner regions of the ejecta, has been interpreted by \cite{Stritzinger2015a} as evidence for a transition from a deflagration to a detonation phase in the pulsational-delayed-detonation (PDD) scenario \citep{hoeflich1995a}. While a deflagration phase of burning is likely required to explain SNe Iax, whether this
is followed by a detonation is still debated as neither pure-deflagration nor delayed-detonation models are able to uniquely reproduce all SN Iax observables.}}

\section{Simulations}
\label{sec:simulations}

In this study, we focus on the previously well-studied model N5def of \citet{fink2014a}, which matches the observational properties of the ``02cx-like'' SN~2005hk well \citep{kromer2013a}. A comprehensive discussion of the explosion model is provided in \citet{fink2014a} and \citet{kromer2013a}.

The explosion was initiated by placing five flame kernels near the centre of the white dwarf in a stochastic way. The small number of ignition sparks induces a strong asymmetry of this ignition configuration causing the ashes to rise buoyantly towards the surface of the white dwarf and to wrap around a core that stays gravitationally bound. About $0.37\, M_\odot$ of material are ejected \citep{kromer2013a, fink2014a}. The evolution is followed up to $100 \, \mathrm{s}$ in the hydrodynamic simulation. A subsequent nucleosynthetic post-processing step determines the chemical composition of the ejected material. It is based on one million tracer particles that record representative thermodynamic trajectories in the explosion simulation and employs a $384$ isotope nuclear network \citep{travaglio2004a, seitenzahl2010a}.

\begin{figure*}
\centering
\includegraphics[width=0.96\textwidth]{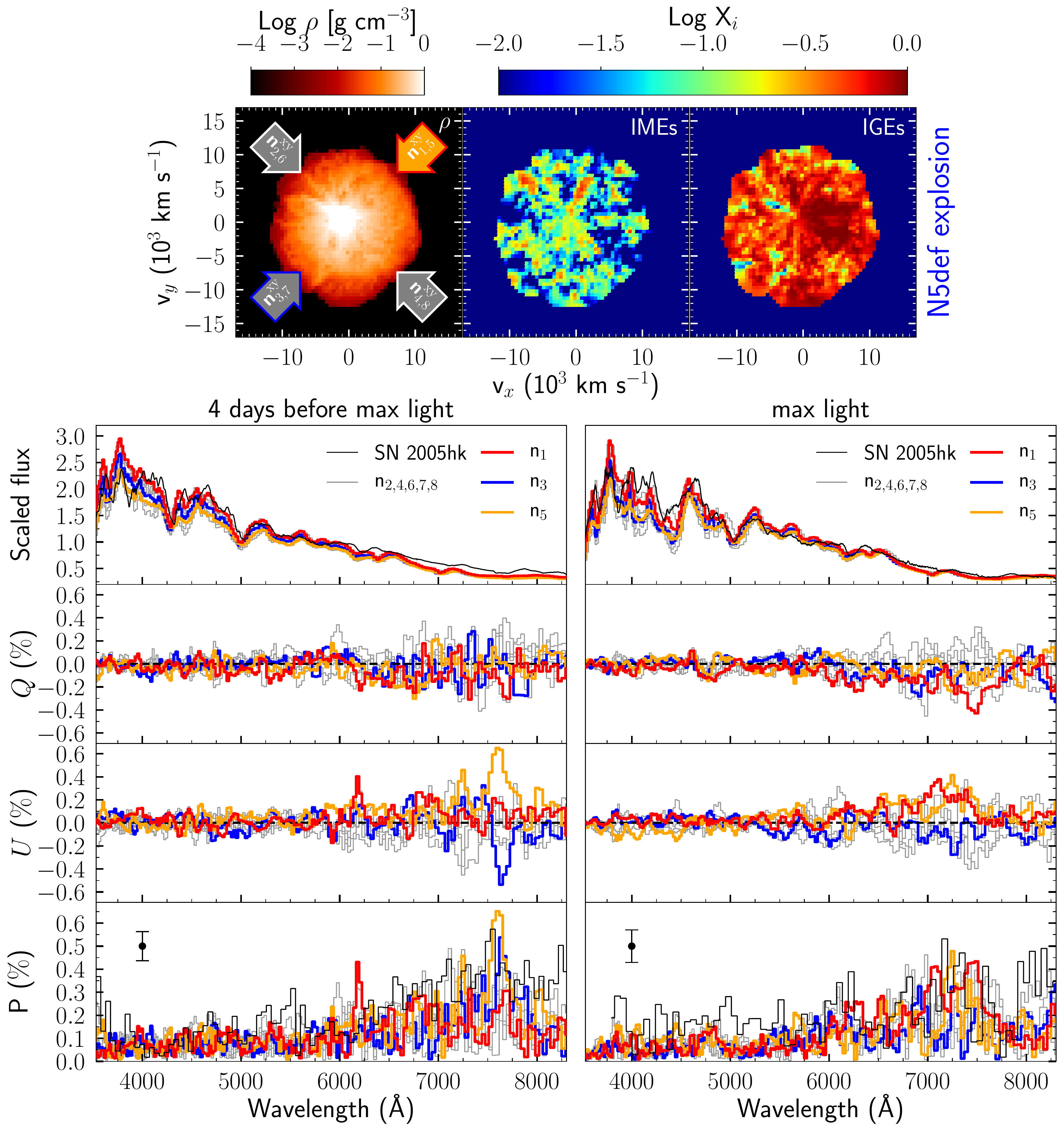}
\caption{Top panels: Ejecta composition for the N5def$^{\,3\mathrm{D}}$ model. Density $\rho$ (left) and mass fraction $X_i$ of IMEs (middle) and IGEs (right) are shown in the $x$--$y$ plane. The $x$--$y$ projected direction of the 8 viewing angles is shown in the density panel. Bottom panels: Flux, $Q$, $U$ and polarisation spectra (from top to bottom) of the N5def$^{\,3\mathrm{D}}$ models are shown 4~days before (left) and at maximum light (right) \revised{for the 8 different orientations ($\textbf{n}_1-\textbf{n}_8$) highlighted in the top left panel}. Flux and polarisation $P$ of SN~2005hk is shown in black for comparison, with polarimetric data from \citet[][bottom right panel]{maund2010a} re-binned of a factor of two to have a similar wavelength resolution ($\Delta\lambda\sim50\,\AA$) and polarisation uncertainties ($\sigma_P\sim0.06$~per~cent, error bars in bottom panels) as those from \citet[][bottom left panel]{chornock2006a}. Both observed and modelled fluxes are normalised at 6000~\AA{} for presentation purposes.}
\label{fig:n5def}
\end{figure*}

As described in detail by \citet{liu2013c}, hydrodynamical simulations of the ejecta--companion interaction for the MS donor $M_\mathrm{ch}$ scenario are performed by using the SPH code \textsc{Stellar GADGET} \citep{Pakmor2012b, Springel2005a}. The MS companion structures at the moment of SN explosion are constructed with 1D detailed binary evolution calculations \citep{han2004a} and then mapped into the SPH code with the HEALPix \citep{gorski2005a} method \citep{Pakmor2012b}.
The angle-averaged 1D ejecta structure of the N5def model is adopted to represent a SN Iax explosion.


The impact is then simulated in 3D for several hours until the SN ejecta--companion star interaction is finished (i.e. when quantities like stripped masses and kick velocities have reached stable values). This interaction creates a conical hole in the supernova debris with an opening angle of about $50^{\circ}$, leaving an asymmetric ejecta structure as seen in the top panels of Figure~\ref{fig:interaction}.  Model\_A in Table 1 of \citet{liu2013c} is used to study the observational imprints of SN ejecta--companion interaction.

We predict synthetic observables for three different models: the 3D N5def model (N5def$^{\,3\mathrm{D}}$), the angle-averaged 1D N5def  model (N5def$^{\,1\mathrm{D}}$), and the companion-interaction (CI) model. All models are re-mapped to a $50^3$ Cartesian grid. Using the Monte Carlo radiative transfer code \textsc{artis} \citep{kromer2009a,bulla2015a}, flux and polarisation spectra are calculated for the N5def$^{\,3\mathrm{D}}$ model to study asymmetries in the explosion, and for the CI model to characterise the signal introduced by the interaction with the companion star. For the N5def$^{\,3\mathrm{D}}$ model, we calculate observables for eight different viewing angles, which are chosen to be representative of each octant in the model: \mbox{$\textbf{n}_{i}=1/\sqrt{3}\,(\pm 1,\pm 1,\pm 1)$}. For the CI model, we restrict the viewing direction to the $x$--$y$ merger plane\footnote{This choice is justified by the fact that the CI model is very close to being axially symmetric about the $x$ axis. This symmetry was verified by predicting spectra for additional viewing angles.} and calculate spectra for five different viewing angles with increasing angular distance $\Delta\Theta$ from the cavity carved out by the companion: $\Delta\Theta=(0,30,45,90,180)^\circ$. As shown in the top panels of Figure~\ref{fig:interaction}, this corresponds to the following observer orientations: $\textbf{m}_1=(-1,0,0)$, $\textbf{m}_2=(-1/2,\sqrt{3}/2,0)$, $\textbf{m}_3=(-1/\sqrt{2},1/\sqrt{2},0)$, $\textbf{m}_4=(0,1,0),$ and $\textbf{m}_5=(1,0,0)$.

Synthetic observables are extracted using the Event-Based Technique (EBT; \citealt{bulla2015a}) in the wavelength range [3500, 10\,000]~\AA{} and between 5 and 16~days after explosion. This time interval covers the $B$-band maximum-light phase $t_B^\mathrm{max}$, occurring 11~days after explosion \citep{kromer2013a}. 
Given that redder regions of the polarisation spectra have typically larger Monte Carlo noise due to lower fluxes, we follow \citet{bulla2015a} and employ a different number of Monte Carlo packets below ($N_\mathrm{q}=4.5\times10^8$) and above ($N_\mathrm{q}=7.5\times10^8$) 5700~\AA. The computed spectra are re-binned in wavelength to achieve a Monte Carlo noise in the polarisation levels of the same order of that in the spectropolarimetric data of SN~2005hk ($\sigma_P\sim$~0.06~per~cent).

\begin{figure*}
\centering
\includegraphics[width=0.95\textwidth]{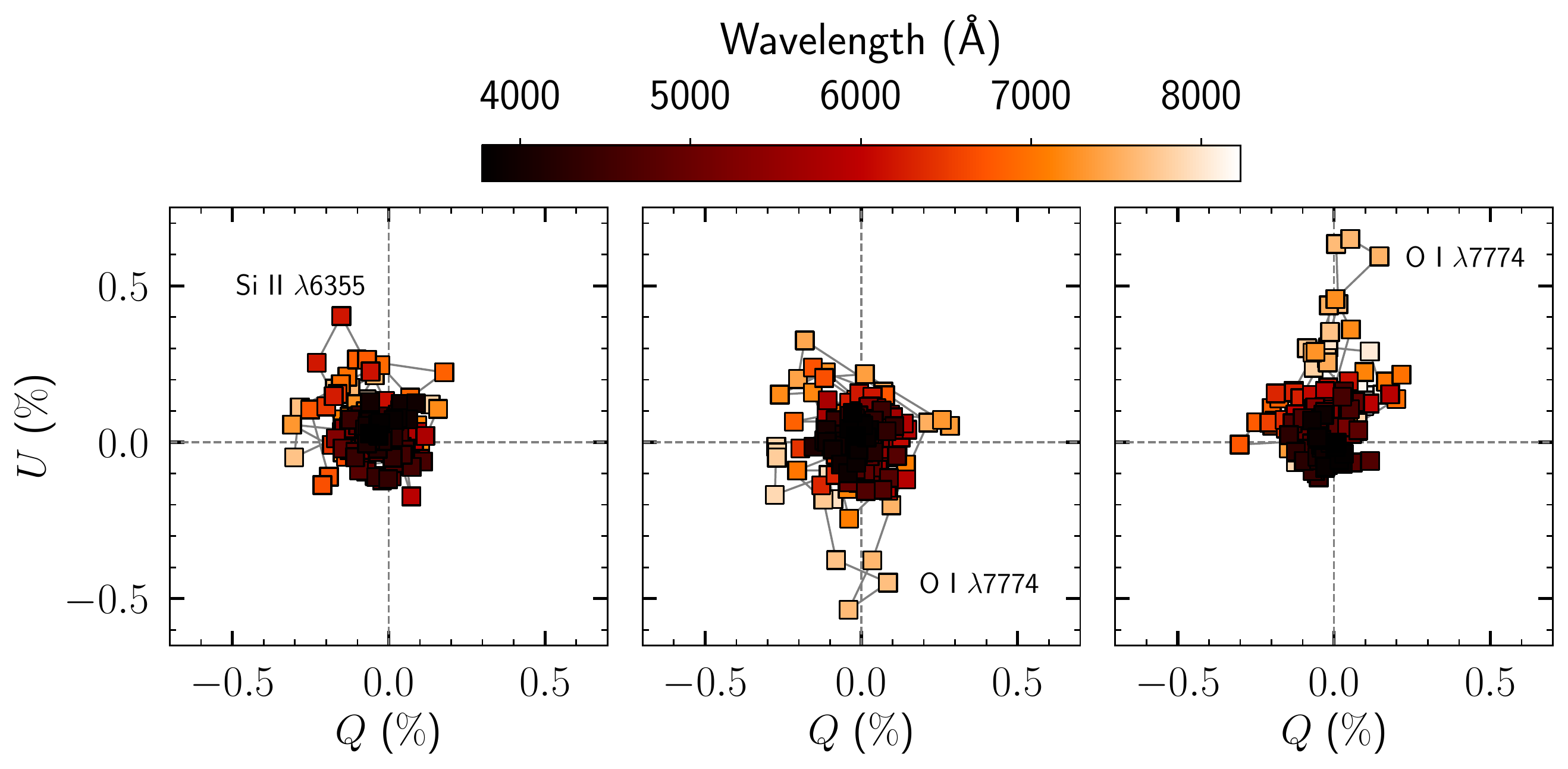}
\caption{\revised{$QU$ plane for the N5def$^{\,3\mathrm{D}}$ model as viewed $4$~days before peak from the \textbf{n}$_1$ (left), \textbf{n}$_3$ (middle) and \textbf{n}$_5$ (right) orientations. These correspond to the 3 spectra highlighted in the left panel of Figure~\ref{fig:n5def} (red, blue, and orange lines, respectively). Loops seen across the Si\,{\sc ii}~$\lambda6355$ (left) and the O\,{\sc i}~$\lambda7774$ (middle and right) lines reflect small-scale asymmetries in the distribution of these elements.} }
\label{fig:quplane}
\end{figure*}

\section{Results}
\label{sec:pol}

In the following, we present flux and polarisation spectra predicted for the N5def$^{\,3\mathrm{D}}$ explosion model (Section~\ref{sec:explosion}) and CI model (Section~\ref{sec:interaction}) and provide comparisons to spectropolarimetric data of SN~2005hk. We focus our discussion on the two epochs with available polarisation spectra, that is, four days before $t_B^\mathrm{max}$ \citep{chornock2006a} and at $t_B^\mathrm{max}$ \citep{maund2010a}. Data are corrected for interstellar polarisation using values reported in \citet{chornock2006a} and \citet{maund2010a}.

\subsection{Signatures from the explosion}
\label{sec:explosion}

Figure~\ref{fig:n5def} shows flux and polarisation spectra calculated for the N5def$^{\,3\mathrm{D}}$ model at four~days before maximum light (left panels) and at maximum light (right panels). The relatively spherical ejecta distribution of the N5def$^{\,3\mathrm{D}}$ model (see top panels of Fig~\ref{fig:n5def}) is reflected in the synthesised spectra. First, the viewing-angle dependence in the flux spectra is modest. In particular, spectra along different orientations are nearly identical redward $\sim$~5000~\AA, while a relatively small difference is predicted between the brightest and faintest orientation at shorter wavelengths (a factor of $\sim$1.5 in flux). These blue wavelengths are dominated by line blanketing from iron-group elements (IGEs, defined as elements from scandium to zinc), thus suggesting that the viewing angle-dependence in this region is caused by small asymmetries in the distribution of IGEs (see top right panel). Second, the predicted $Q$ and $U$ polarisation signals are rather small; the overall levels are $\lesssim$~0.2~per cent at both epochs and across the wavelength range investigated. Polarisation across individual spectral features is larger but still relatively modest; the strongest signals are seen four~days before peak across the Si\,{\sc ii}~$\lambda6355$ line at $\sim$6200~\AA{} ($P\sim0.4$~per cent along \textbf{n}$_1$, red line) and the O\,{\sc i}~$\lambda7774$ line at $\sim$7500~\AA{} ($P\sim0.6$~per cent along \textbf{n}$_3$, blue line, and \textbf{n}$_5$, orange line). \revised{As shown in Figure~\ref{fig:quplane}, distinctive loops are seen across these lines in the $QU$ plane, which is suggestive of small-scale asymmetries in the distribution of silicon and oxygen within the ejecta (see e.g. \citealt{wang2008a} and \citealt{Bulla2017PhD} for detailed reviews of loops in the $QU$ planes of SNe). \revisedtwo{The presence of loops provides evidence for departures from a simple bi-axial (e.g. ellipsoidal; see \citealt{Stevance2019}) configuration of the ejecta and is a natural consequence of mixing in a deflagration phase of burning, which is common to our model and the PDD models invoked by \cite{Stritzinger2015a}.}}

\begin{figure*}
\centering
\includegraphics[width=0.96\textwidth]{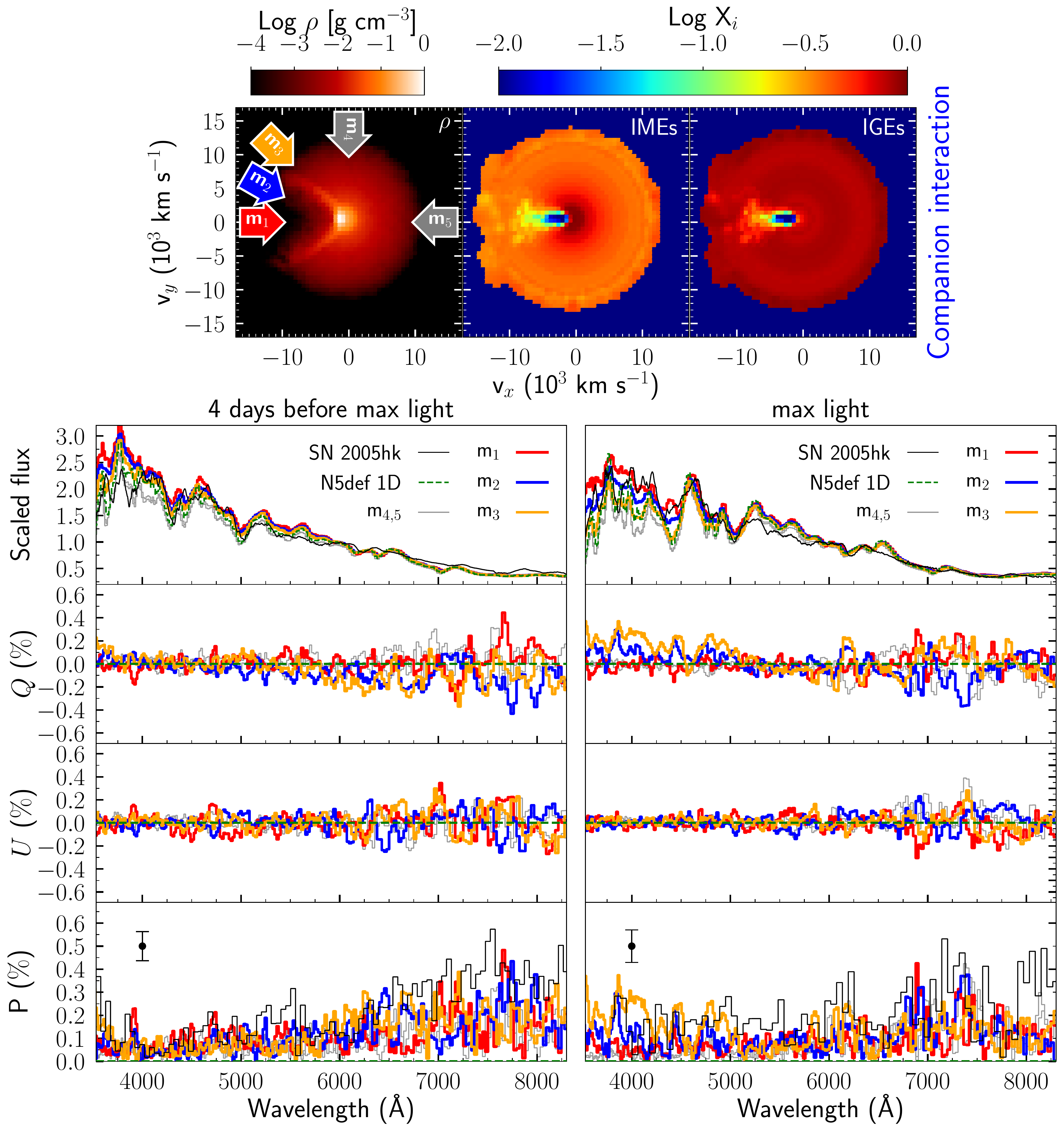}
\caption{Same as Figure~\ref{fig:n5def} but for the CI model \revised{viewed from the 5 different orientations ($\textbf{m}_1-\textbf{m}_5$) highlighted in the top left panel}. The green dashed lines in the top panels are spectra calculated for the N5def$^{\,1\mathrm{D}}$ model, thus referring to the case with no interaction. The N5def$^{\,1\mathrm{D}}$ model is spherically symmetric by construction and thus no polarisation is expected by definition (see horizontal dashed lines in bottom panels).}
\label{fig:interaction}
\end{figure*}

Observed flux and polarisation spectra of SN~2005hk are shown in black in Figure~\ref{fig:n5def}. As already reported by \citet{kromer2013a}, we find a good agreement at both epochs between the predicted flux spectra and those observed in SN~2005hk. A reasonable agreement is also seen in the polarisation signal, and both modelled and observed spectra are characterised by an overall low polarisation level and the absence of strong polarisation features. Nevertheless, we note some differences. In particular, the polarisation signal shortward of $\sim$~5000~\AA{} increases from the first to second epoch in SN~2005hk, while no significant change is predicted in the N5def$^{\,3\mathrm{D}}$ model. As a consequence, the N5def$^{\,3\mathrm{D}}$ model predicts somewhat smaller signals at peak brightness compared to those observed. We come back to this discrepancy in Section~\ref{sec:interaction}.

\subsection{Signatures from the ejecta--companion interaction}
\label{sec:interaction}

Figure~\ref{fig:interaction} shows synthetic observables predicted for the CI model four~days before peak (left panels) and at peak (right panel). The green dashed lines in the top panels show spectra for the N5def$^{\,1\mathrm{D}}$ model, which are plotted to gauge the impact of the ejecta--companion interaction on the spectra. This comparison highlights how the interaction affects the spectra only below $\sim$~5000~\AA, and there are no clear signatures at longer wavelengths.

Similar to the N5def$^{\,3\mathrm{D}}$ model, the viewing-angle dependence in the CI model is confined to IGE-dominated regions below $\sim$~5000~\AA, while almost identical spectra are predicted at longer wavelengths for different observer orientations. The angular dependence can be understood as a direct consequence of the location of the different observers with respect to the cavity carved out by the companion star. As shown in the top left panel of Figure~\ref{fig:interaction}, photons can more easily escape the system along \textbf{m}$_1$ compared to orientations away from the cavity (\textbf{m}$_2$ to \textbf{m}$_5$). The brightest spectrum is therefore associated with the \textbf{m}$_1$ orientation (red line). An increase in optical depth is found when moving the observer away from the cavity, thus leading to increasingly fainter spectra from \textbf{m}$_2$ ($\Delta\Theta=30^\circ$) to \textbf{m}$_4$ ($\Delta\Theta=90^\circ$). Finally, the optical depth in the \textbf{m}$_5$ direction ($\Delta\Theta=180^\circ$) is relatively similar to that in the \textbf{m}$_4$ direction, and thus the corresponding spectra are rather similar.

The viewing-angle dependence of the polarisation spectra follows the ejecta distribution of the CI model. In particular, the ejecta are relatively symmetric in projection when viewed from \textbf{m}$_1$, that is when looking directly down the cavity ($\Delta\Theta=0^\circ$). The corresponding polarisation signal is therefore very close to zero. Similarly low polarisation levels are predicted along \textbf{m}$_4$ and \textbf{m}$_5$. In contrast, the presence of the cavity breaks the projected symmetry in the \textbf{m}$_2$ and \textbf{m}$_3$ directions, thus producing stronger polarisation signals along these two viewing angles. In particular, the interaction with the companion star produces a maximum-light polarisation signal of $P\sim$~0.2~per cent below $\sim$~5000~\AA, a distinctive feature that is not seen in the N5def$^{\,3\mathrm{D}}$ explosion model. This reflects the break of symmetry introduced by the cavity, whose effect is stronger at bluer wavelengths where the IGEs are dominant.

Polarisation predicted for the CI model is in reasonable agreement with that observed for SN~2005hk at both epochs. Similar to the N5def$^{\,3\mathrm{D}}$, the overall low polarisation signals and absence of strong polarisation features are consistent with spectropolarimetric data of SN~2005hk.
We note, however, a significant difference. While the N5def$^{\,3\mathrm{D}}$ model was not polarised enough to explain the polarisation signal of $P\sim$~0.2~per cent detected for SN~2005hk at maximum light and below $\sim$~5000~\AA, a better match to data is found for the CI model seen from \textbf{m}$_2$ and \textbf{m}$_3$. This suggests that the polarisation signal observed at blue wavelengths in SN~2005hk might stem from the interaction between the SN ejecta and the companion star. More polarimetric data of SNe~Iax is needed to confirm this interpretation.



\section{Discussion and conclusions}
\label{sec:conclusions}

In this \revisedtwo{paper}, we carry out radiative transfer calculations and predict signatures from the deflagration of a $M_\mathrm{ch}$ white dwarf and the subsequent interaction of the ejecta with a companion star. To investigate signatures arising from the explosion, we first study the 3D deflagration model N5def$^{\,3\mathrm{D}}$ \citep{kromer2013a,fink2014a} and predict flux and polarisation spectra for eight different orientations. The viewing-angle dependence of the flux spectra and the overall polarisation signals are rather small, owing to the relatively spherical ejecta distribution of the N5def$^{\,3\mathrm{D}}$ model and to the dominant contribution of IGEs that tend to depolarise the escaping radiation. 

To estimate the impact of the ejecta--companion interaction, we use simulations from \citet{liu2013c}, where an angle-averaged 1D ejecta of the N5def$^{\,3\mathrm{D}}$ model (N5def$^{\,1\mathrm{D}}$) collides with a MS companion star, and predict synthetic observables for five different viewing angles. We note that a He star is more compact but closer to the white dwarf compared to a MS star, thus creating a cavity with a similar opening angle. We therefore expect similar results in the He star case, a possibility that we plan to address in a future study. Overall, the imprint of the ejecta--companion interaction on both flux and polarisation spectra is only modest and restricted to wavelengths below $\sim$~5000~\AA. The brightest orientation is found when looking down the cavity carved out by the interaction, while increasingly fainter orientations are predicted when moving away from the cavity. This trend is in qualitative agreement with predictions from \citet[][see their figure 3]{kasen2004a}. 

The polarisation level in the ejecta--companion interaction model is maximum for orientations 30$^\circ$ and 45$^\circ$ away from the cavity; negligible signals are predicted when the system is viewed down the cavity or 90$^\circ$ and 180$^\circ$ away from it. This is in contrast with predictions from \citet{kasen2004a}, who found the continuum polarisation level to peak around 90$^\circ$ (see their figures 7 and 10). Given that the choice of the cavity opening angle is similar to that found in our simulations, we interpret the discrepancy as due to the different composition structures and particularly IGE distributions. \citet{kasen2004a} assumed a layered structure adopting the so-called W7 model for normal SNe~Ia \citep{nomoto1984a}, where IGEs are confined to the inner regions. In contrast, our interaction model that is based on a 3D explosion simulation is characterised by IGEs that extend throughout the whole ejecta including the cavity created by the interaction. Strong line opacities from IGEs tend to destroy most of the polarisation signal created by electron scattering, thus likely explaining the smaller polarisation signals in our compared to the \citet{kasen2004a} simulations.

Guided by the increasing evidence suggesting that SNe~Iax result from deflagrations of $M_\mathrm{ch}$ white dwarfs in binary systems, we compare our simulations to the only spectropolarimetric data available for a SN~Iax, namely SN~2005hk \citep{chornock2006a,maund2010a}. The observed low polarisation signals and lack of strong polarisation features in SN~2005hk are well reproduced by our models. Interestingly, the ejecta--companion interaction model provides a better agreement to data at maximum light and at blue ($\lesssim$~5000~\AA) wavelengths, where the predicted level of $P\sim0.2$~per~cent matches that observed in SN~2005hk. This level is hard to achieve from the explosion itself given the overall spherical symmetry of the N5def$^{\,3\mathrm{D}}$ model and most importantly the strong depolarising line blanketing at these short wavelengths (although we anticipate exploring more aspherical deflagration models in a future work). \revised{We note that a rise in polarisation from shorter to longer optical wavelengths is not a unique prediction of our deflagration model, with the same effect observed for two sub-luminous SNe~Ia (SN~1998by; \citealt{howell2001a}, and SN~2005ke, \citealt{patat2012a}) and interpreted as a line blanketing effect in the framework of fast-rotating white dwarfs or white dwarf mergers \citep{patat2012a}.}

\revised{Given that our comparison is limited to one object, it is hard to assess whether or not our results can be generalised to the whole SN Iax class.} We thus encourage further polarimetric observations of SNe~Iax covering short optical wavelengths, for which an ejecta--companion interaction signature should be detectable in $\sim6\%$ of the cases (assuming orientations with $30^\circ\lesssim\Delta\Theta\lesssim60^\circ$ from the cavity).

In this pilot study, we chose to disentangle the asymmetries introduced by the explosion with those originated by the ejecta--companion interaction. In the future, we will combine these two effects and carry out self-consistent calculations in which the full 3D ejecta structure predicted by deflagration models collides with the companion star. We also plan to make predictions for different scenarios, including different companion stars (e.g. He star) and/or using different explosion models more tailored to normal SNe~Ia.

\section*{Acknowledgements}

\revised{We thank the anonymous reviewer for his/her valuable comments. We are grateful to Ryan Chornock and Justyn Maund for sharing the spectropolarimetric data of SN\,2005hk.} This work used the DiRAC Complexity system, operated by the University of Leicester IT Services, which forms part of the STFC DiRAC HPC Facility (\url{www.dirac.ac.uk}). This equipment is funded by BIS National E-Infrastructure capital grant ST/K000373/1 and STFC DiRAC Operations grant ST/K0003259/1. DiRAC is part of the National E-Infrastructure. This research was supported by the Partner Time Allocation (Australian National University), the National Computational Merit Allocation and the Flagship Allocation Schemes of the NCI National Facility at the Australian National University. Parts of this research were conducted by the Australian Research Council Centre of Excellence for All-sky Astrophysics (CAASTRO), through project number CE110001020. The authors gratefully acknowledge the Gauss Centre for Supercomputing (GCS) for providing computing time through the John von Neumann Institute for Computing (NIC) on the GCS share of the supercomputer JUQUEEN \citep{Stephan2015} at J\"ulich Supercomputing Centre (JSC). GCS is the alliance of the three national supercomputing centres HLRS (Universit\"at Stuttgart), JSC (Forschungszentrum J\"ulich), and LRZ (Bayerische Akademie der Wissenschaften), funded by the German Federal Ministry of Education and Research (BMBF) and the German State Ministries for Research of Baden-W\"urttemberg (MWK), Bayern (StMWFK) and Nordrhein-Westfalen (MIWF). ZWL is supported by the National Natural Science Foundation of China (NSFC, No. 11873016) and the 100 Talents Programme of the Chinese Academy of Sciences. The work of FKR is supported by the Klaus Tschira Foundation and through the Sonderforschungsbereich SFB 881 ``The Milky Way System'' of the German Research Foundation (DFG). SAS acknowledges support from STFC through grant ST/P000312/1. IRS was supported by the Australian Research Council through grant No.~FT160100028.

\bibliographystyle{aa}
\bibliography{astrofritz}

\end{document}